# Scalar damage model for concrete without explicit evolution law

J. Ožbolt
*Institute of Construction Materials, University of Stuttgart*
*Pfaffenwaldring 4, 70569 Germany*
*Email: ozbolt@iwb.uni-stuttgart.de*

S. Ananiev
*Institute of Lightweight Structures and Conceptual Design, University of Stuttgart*
*Pfaffenwaldring 7, 70569 Germany*
*Email: sergey.ananiev@po.uni-stuttgart.de*

ABSTRACT: Based on the fact that for an isotropic material model the elastic predictor and the projected stress tensors have the same eigenvectors, it is shown that the scalar damage can be obtained directly from the projection algorithm. This eliminates the difficulty of a proper definition of equivalent strain which serves as a driving force for evolution of damage in concrete. Moreover, if eigenvectors are known it is not more necessary to use invariants of the stress tensor for the formulation of loading surface. In the present model the loading surface is represented in the polynomial form. This has two advantages: (i) it automatically fulfils $C_1$ continuity and (ii) plane stress formulation is achieved by simply setting the third stress to zero. The performance of the model is illustrated on example of a mixed-mode fracture of concrete. It is shown that for the present example the model prediction strongly depends on the choice of the stress degradation law.

## 1 INTRODUCTION

Damage and plasticity theories are well established theoretical frameworks for macroscopic modelling of materials. One difficulty in the modelling of concrete like materials is due to the fact that such materials exhibit highly non-symmetrical behaviour in tension and compression. Consequently, damage formulations based on the equivalent strain concept are difficult to extend to concrete.

The problem can be partially solved by taking only the positive (tensile) part of the strain tensor as a driving force of damage (Mazars, 1986). This seems reasonable, however, the absence of the loading surface leads to mismatching of elastic and inelastic ranges with a consequence that, for instance, under biaxial compression damage takes place too early.

In plasticity-like formulation (see for instance Meschke et al., 1998) the loading surface is a starting point. Degradation of the elastic stiffness tensor is related to the derivates of the loading surface following the postulate of maximum energy dissipation functional or, for softening materials, using the postulate of its stationarity. The problem here is a validity of this postulate for concrete like materials. Even under biaxial compression the inelastic deformations do not have the same nature as plastic deformations in metals. There is no crystal slip simply because the material does not have a crystalline structure. Therefore, it is not surprising that the non-associated flow rule, which is prohibited by the above mentioned postulate, often produces better results.

Here, a scalar damage model for concrete is formulated based on a procedure that is formally the same as the plasticity-like formulation. As well known, the main purpose of the flow rule is a determination of six equations for six unknown components of plastic stain tensor. Additional unknown, the rate of plastic strains, is obtained from the consistency condition, i.e. during inelastic loading the stress state can not leave the loading surface.

Formally, one can say that if there are no plastic strains (elastic damage), there is no need for the flow rule. If damage is defined with only one unknown (scalar damage), than it can be obtained from the consistency condition. The assumption of scalar damage has a consequence that the material remains isotropic for the entire load history and that the current damage can be determined by scaling of elastic stress state back to the loading surface. The loading surface itself is parameterized using the actual values (strengths) of the uniaxial tensile and uniaxial compressive stresses. Since the loading surface is formulated in the principal stress space, their dependences on damage variable can be taken directly from experiments.

## 2 FORMULATION OF THE MODEL

### 2.1 Loading surface

The idea that inelastic (plastic) deformations initiate after the equivalent stress reaches a certain level (yield stress or strength of the material) can be traced back to the work of von Mises. It is reasonable to assume that for isotropic material the equivalent stress is independent of the choice of the coordinate system. This is the reason why the expression for equivalent stress was formulated using invariants of the stress tensor. Unfortunately, for concrete like materials it is not a simple task to formulate loading surface in a closed form. One possible way is a formulation in Haigh-Westergaard coordinates as it has been done, for example, by Etse (1992). The loading surface is continuously differentiable, however, it has a relatively large number of parameters, which complicates the formulation. A more serious disadvantage of this model, which hold for every plasticity model, is the fact that the tension softening is understood as accumulation of plastic strains, which is physically not correct.

Another possibility is a multi-surface formulation where more loading surfaces with a relatively low number of parameters per surface are used. The disadvantage here is that the resulting surface is non-continuously differentiable. This is from a numerical point of view not a problem since the corresponding value of plastic multiplier can be obtained using Koiter's rule. However, from the physical point of view it is not acceptable that a infinitely small rotation of stress tensor causes a finite rotation of plastic strain tensor.

When the loading surfaces are expressed in terms of stress invariants than it is difficult to connect them smoothly. Therefore, to avoid the problem of the discontinuity when skipping from one loading surface to another, in the present model a polynomial representation of the entire loading surface is employed. The components of the polynomial expression are principal stresses. The justification of not using invariants of stress tensor can be found in Simo (1992). He has shown that for isotropic material models the stress projection algorithm does not change the eigenvectors and if they are known it is not longer necessary to use stress invariants.

An additional advantage of such a formulation is that the reduction to the plane stress state is achieved by simple setting of one principal stress to zero. This removes the need for some special techniques like subiterations (Feenstra, 1993) or a projection matrix (Simo and Taylor,1986).

For reasons of simplicity, in the following the formulation of the model is restricted to the two-dimensional state of stresses. The proposed loading surface $F$, which is a fourth order polyinom with seven unknown coefficients, reads:

$$F(\sigma_I, \sigma_{II}) \equiv 0$$
$$a_1(\sigma_I^2 \cdot \sigma_{II}^2) + a_2(\sigma_I^3 + \sigma_{II}^3) + a_3(\sigma_I^2 \cdot \sigma_{II} + \sigma_{II}^2 \cdot \sigma_I) + \quad (1)$$
$$+ a_4(\sigma_I^2 + \sigma_{II}^2) + a_5 \sigma_I \cdot \sigma_{II} + a_6(\sigma_I + \sigma_{II}) + a_7 = 0$$

where $\sigma_I$ and $\sigma_{II}$ are principal stresses and $a_1$ to $a_7$ are unknown coefficients. Seven coefficients are sufficient to realistically describe the loading function in two-dimensional stress space. They are obtained from the following seven conditions:

- The biaxial compressive strength $k_1 f_C$ is larger than the uniaxial compressive strength $f_C$:

$$F(-k_1 \cdot |f_C|, -k_1 \cdot |f_C|) = 0 \quad (i)$$

- The maximal compressive strength $k_2 f_C$ is achieved when the stress in one direction is two times larger than in the other direction. The maximal compressive strength is larger than the biaxial compressive strength ($k_2 > k_1$):

$$F(-0.5 k_2 \cdot |f_C|, -k_2 \cdot |f_C|) = 0 \quad (ii)$$

$$\frac{\partial F}{\partial \sigma_I}(-0.5 k_2 \cdot |f_C|, -k_2 \cdot |f_C|) = 0 \quad (iii)$$

- The uniaxial compressive strength is $f_C$:

$$F(0, -|f_C|) = 0 \quad (iv)$$

- The uniaxial tensile strength $f_T$:

$$F(f_T, 0) = 0 \quad (v)$$

- The Rankine tension cut-off criteria reads:

$$\frac{\partial F}{\partial \sigma_I}(f_T, 0) = 1 \quad (vi)$$

$$\frac{\partial F}{\partial \sigma_{II}}(f_T, 0) = 0 \quad (vii)$$

The above specified seven boundary conditions allow us to find seven constants from (1). The resulting loading surface is shown in Figure 1. It is similar to the curve obtained from experiments. At the points A and B the loading surface is automatically $C_1$ continuous because of the symmetry of the polynom with respect to the principal stresses.

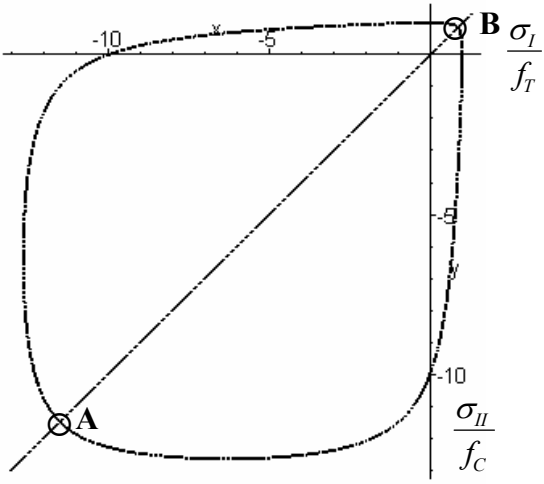

Figure 1. Loading surface defined as a fourth order polynom.

## 2.2 Softening rules

Similar to the idea of Feenstra (1993), the softening rule is taken directly from the uniaxial experiment. The softening rules are understood here as dependence between tensile/compressive stress and scalar damage. The objectivity with respect to the finite element size is accounted for by scaling of fracture energy $G_F$ to the average element size, i.e. the crack band approach is employed (Bažant and Oh, 1983). In the following the description of several different softening curves that are implemented into the model is discussed.

### 2.2.1 Tension: linear degradation

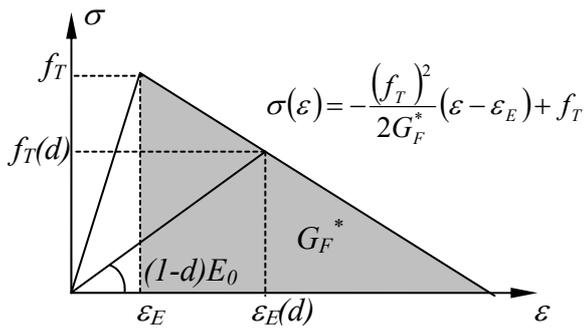

Figure 2. Linear degradation of tensile stress.

As a first approximation let's assume a linear degradation of tensile stress. The dependence between tensile stress $f_T$ and scalar damage $d$ can be obtained by solving an equation which states that for a given stress one has to get the same elastic strain from the damage formulation and from the degradation equation:

$$\frac{f(d)}{(1-d)E_0} = \left( f_T + \frac{f_T^2 \varepsilon_E}{2G_F^*} - f(d) \right) \frac{2G_F^*}{f_T^2} \quad (2)$$

where $G_F^*$ is the area under the softening stress-strain curve that is related to concrete fracture energy $G_F$ and element size $h$ (crack band approach) and $E_0$ is the Young's modulus. From (2) the desired dependence reads:

$$f(d) = (1-d)f_T \frac{2G_F^* E_0 + f_T^2}{2G_F^*(1-d)E_0 + f_T^2} \quad (3)$$

It is obvious that the softening law is not a linear function of damage.

### 2.2.2 Tension: discontinuous linear degradation

To study the influence of a sharp discontinuity in the stress-strain curve after onset of cracking, the stress-strain softening curve shown in Figure 3 is implemented into the model.

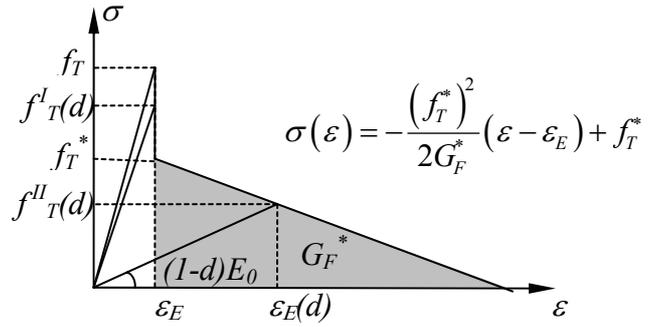

Figure 3. Discontinuous linear degradation of tensile stress.

The discontinuity of the degradation law for tensile stress does not mean that the softening law as a function of damage is discontinuous as well. It reads:

$$f(d) = (1-d)f_T \quad (\varepsilon = \varepsilon_E)$$
$$f(d) = (1-d)f_T^* \frac{2G_F^* E_0 + (f_T^*)^2}{2G_F^*(1-d)E_0 + (f_T^*)^2} \quad (\varepsilon > \varepsilon_E) \quad (4)$$

### 2.2.3 Tension: exponential degradation

The exponential degradation law is frequently used in the literature, (Feenstra, 1993; Mazar, 1986; Pivonka et al., 2002). Unfortunately, it does not allow a closed form formulation of the softening law for $f(d)$ which somewhat complicates the formulation.

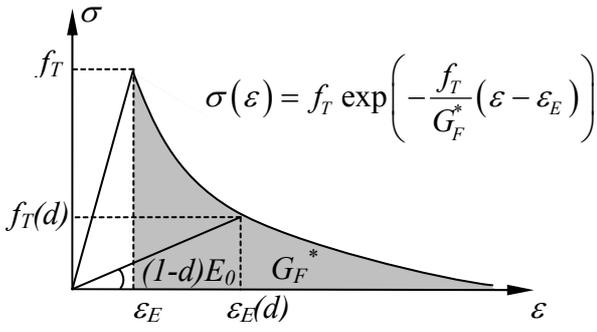

Figure 4. Exponential degradation of tensile stress.

The desired softening law is now expressed in implicit form:

$$\Theta(f,d) = 0$$

$$f(d) + (1-d)E_0 \frac{G_F^*}{f_T} \ln(f(d)) - $$

$$-(1-d)\left(E_0 \frac{G_F^*}{f_T} \ln(f_T) - f_T\right) = 0 \quad (5)$$

$$\frac{\partial \Theta}{\partial f}(f,d) = 1 + \frac{1}{f}(1-d)E_0 \frac{G_F^*}{f_T}$$

An analytical solution of (5) is possible using Lambert's function, but here the solution is obtained numerically using standard Newton-Rathson method:

$$f(d^*) = f_0 - \sum_i \frac{\Theta(f_{i-1},d^*)}{\frac{\partial \Theta}{\partial f}(f_{i-1},d^*)} \quad (6)$$

where $d^*$ is a given value of damage for which the corresponding value of tensile stress has to be found. Equation (6) is a general solution and it makes possible to use any form of degradation law.

### 2.2.4 Compression: parabolic degradation

Damage approach describes the behaviour of concrete in tension reasonably well. However, from experimental evidence it is obvious that for dominant compressive load a plasticity plays an important role as well. Principally, a realistic model for concrete has to take into account both phenomena, i.e. damage and plasticity. When a single loading surface is used, then it is necessary to separate somehow these inelastic processes. Meschke et al. (1998) proposed to use a simple scalar multiplicator, which controls the proportion between dissipated energy during plastic and damage deformations.

In the present model, two separate and non-overlapping loading surfaces are proposed. Non-overlapping in this context means that the plastic strains are accumulated only during hardening and that the softening is modelled exclusively by scalar damage. Moreover, it is assumed that the hardening affects only compressive stress (see also Feenstra, 1993). This allows independent consideration of plasticity and damage. The proposed stress-strain relationship for uniaxial compression is illustrated in Figure 5. In the following, no hardening is accounted for ($\varepsilon^p = 0$) since the plasticity formulation is out of the scope of the present paper.

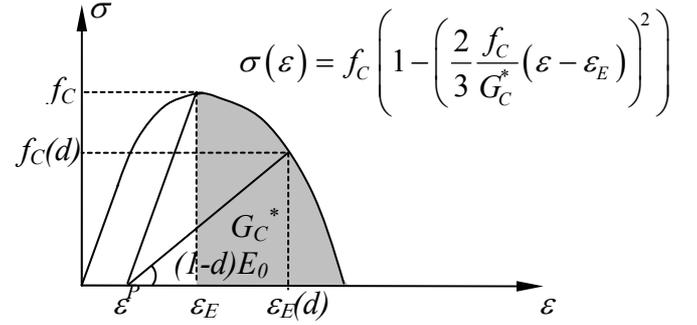

Figure 5. Parabolic degradation of compressive stress.

The desired softening law is obtained as a first (positive) root of the quadratic equation:

$$f(d) = -\frac{1}{2}A(d) + \frac{1}{2}\sqrt{(A(d))^2 - 4B(d)}, \quad (7)$$

where,

$$A(d) = -2(1-d)f_C + \frac{((1-d)E_0)^2}{f_C}\left(\frac{3}{2}\frac{G_C^*}{f_C}\right)^2,$$

$$B(d) = (1-d)^2\left(f_C^2 - \left(\frac{3}{2}\frac{G_C^*}{f_C}E_0\right)^2\right). \quad (8)$$

## 3 RESOLVING EQUATION

For simplicity the formulation of the model is illustrated in Figure 6 assuming a Tresca loading surface. The actual value of damage is obtained from scaling of the elastic stress state back to the loading surface, but not to the initial one. The loading surface is parameterized using uniaxial tensile and compressive stresses, which depend on the actual level of damage. That means, the failure surface "shrinks" with increase of damage (see Figure 6).

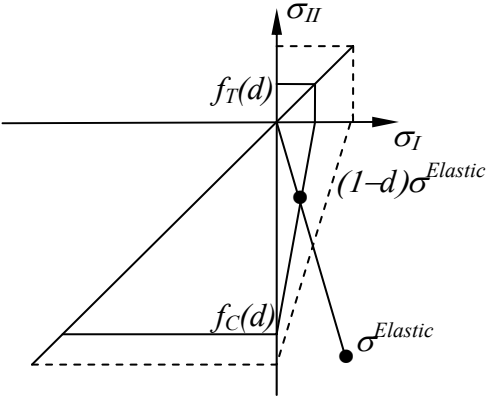

Figure 6. Projection of elastic stress state back to the loading surface.

Such scaling can be understood as a simplified form of a backward Euler integration scheme, which can be reduced to one scalar equation regarding the plastic multiplier. For example, the consistency condition for the case shown in Figure 6 reads:

$$(1-d)\left(\frac{\sigma_I^{Elastic}}{f_T(d)}+\frac{\sigma_{II}^{Elastic}}{f_C(d)}\right)=1 \quad (9)$$

In special case, where both softening laws have linear form (3), it is possible to solve (9) analytically. The solution reads:

$$d(\sigma_I,\sigma_{II})=\frac{CD(\sigma_I f_C - f_T f_C + \sigma_{II} f_T)}{\sigma_I(C-f_T^2)f_C D+\sigma_{II}(D-f_C^2)f_T C} \quad (10)$$

where,

$$C=2G_F^* E_0 + f_T^2,$$
$$D=2G_C^* E_0 + f_C^2. \quad (11)$$

Actually, (10) represents a damage evolution law, but it is formulated in completely different way as one usually does, i.e. without the equivalent strain concept.

## 4 ALGORITHMIC TANGENT OPERATOR

In the plasticity formulation the continuous tangent operator does not provide a quadratic convergence of the Newton-Rathson method during equilibrium iterations (Simo and Taylor, 1986). One has to linearize the stress state equations after projection to the loading surface.

In the presented formulation the scalar damage does not result from the integration of a certain flow rule. Consequently, there is no continuous tangent operator. The only possible tangent operator is a algorithmic one derived from the projected stress state. A general expression for scalar damage can be written as follows:

$$\boldsymbol{\sigma}=(1-d)\boldsymbol{D}^{el}:\boldsymbol{\varepsilon} \quad (12)$$

Differentiation of (12) gives the tangent operator:

$$\frac{\partial\boldsymbol{\sigma}}{\partial\boldsymbol{\varepsilon}}=(1-d)\boldsymbol{D}^{el}-\frac{\partial d}{\partial\boldsymbol{\varepsilon}}\otimes\boldsymbol{\sigma}^{Elastic} \quad (13)$$

The partial derivates of damage required in (13) can be obtained from (10). During unloading the tangent operator automatically reduces to the secant one because damage does not change, i.e. it is independent of strains. It is interesting to observe that although a relatively simple isotropic scalar damage model is considered, the tangent stiffness operator (13) is in general non-symmetric.

## 5 NUMERICAL EXAMPLE

Recent numerical studies showed that relatively complex mixed-mode failure of concrete can be objectively simulated using smeared fracture finite element analysis only if a sophisticated material model is used (Ožbolt and Reinhardt, 2003). The studies were carried out on the Double-Edge-Notched specimen tested by Nooru-Mohamed (1992) (see also Pivonka et al., 2002). It has been observed that the material model plays the most important role. To investigate the performance of the presented scalar damage model the numerical analysis is performed on the same example.

The geometry and the test set-up are shown in Figure 7.

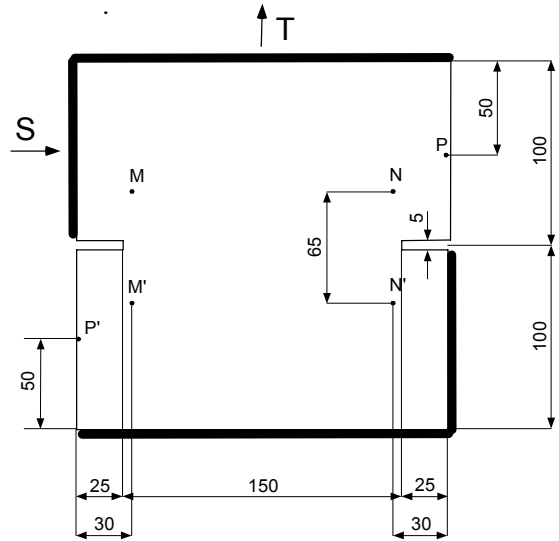

Figure 7. Geometry of the Nooru-Mohamed test specimen.

The specimen was first loaded by shear load $S$. Subsequently, at a constant shear load of $S = 10$ kN, the vertical tensile load $T$ was applied up to failure. The load control procedure was used by moving the upper loading platens in horizontal and vertical direction, respectively. The rotation of the loading platens was restricted. During the application of the horizontal load $S$, the vertical load was kept at zero ($T = 0$). Upon subsequent tensile loading the shear force was kept constant. The bottom (support) platens were fixed and, the same as the upper (loading) platens, glued to the surface of the specimen. The finite element discretization is performed by plane stress quadrilateral finite elements with four integration points. The material properties are taken as: Young's modulus $E_C = 32800$ MPa, Poisson's ratio $v = 0.2$, tensile strength $f_T = 3.0$ MPa, uniaxial compressive strength $f_C = 38.4$ MPa and concrete fracture energy $G_F = 0.11$ N/mm.

For the present example the compressive properties of concrete are of minor importance. For this reason only the influence of the tensile softening laws was investigated. The experimentally observed crack pattern is shown in Figure 8. As can be seen the crack pattern consists of two separate cracks.

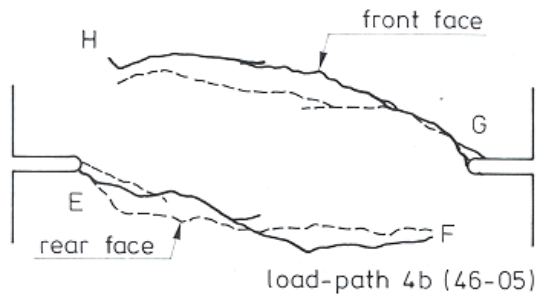

Figure 8. Crack pattern observed in the experiment.

The first softening law that was linear. The fracture energy $G_F$ was scaled to the width of the crackband $h = 5$ mm (element size). The observed crack pattern is shown in Figure 9.

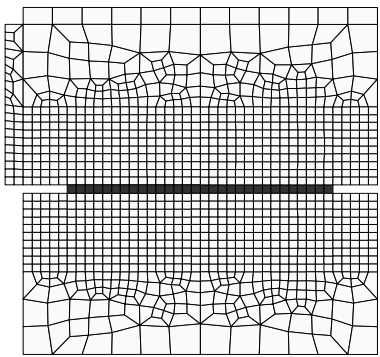

Figure 9. Crack pattern obtained by employing a linear softening law.

As it can be seen, the use of the linear softening law results in an unrealistic crack pattern, i.e. the crack plotted in terms of maximal principal strains propagates along straight line and it is not curved as in the experiment. The corresponding load-displacement curve (Figure 12) indicates an overestimation of the peak resistance obtained in the experiment.

By introducing discontinuity in the degradation law for tensile stress ($f_T^*/f_T = 0.8$, see Figure 3) the crack path becomes more realistic and the calculated load-displacement curve agrees better with the experiment.

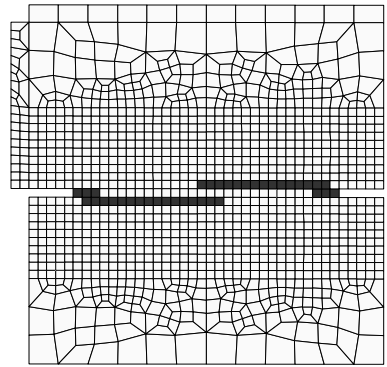

Figure 10. Crack pattern obtained by the use of discontinuous linear degradation.

The third simulation was performed by employing a linear degradation law with extremely low fracture energy $G_F$. It was reduced by a factor of 150. The concrete in this case is thus assumed to be nearly a brittle material. This explains a steep decrease of the softening branch of the load-displacement curve and consequently significant underestimation of the ultimate load (see Figure 12). However, as can be seen from Figure 11 the calculated crack pattern agrees well with the experimental one.

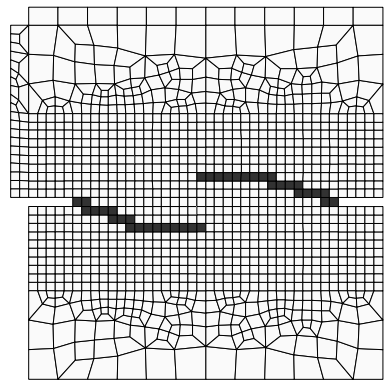

Figure 11. Crack pattern obtained assuming a brittle material.

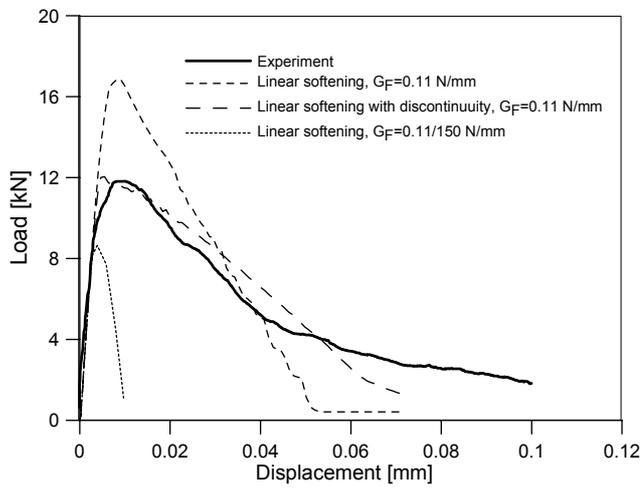

Figure 12. Load-displacement curves obtained in all three numerical simulations as well as the experimentally obtained curve.

To get an insight in the behaviour of the model it is useful to draw the softening laws based on the linear degradation of tensile stress (see Figure 2) for different values of the crack band width.

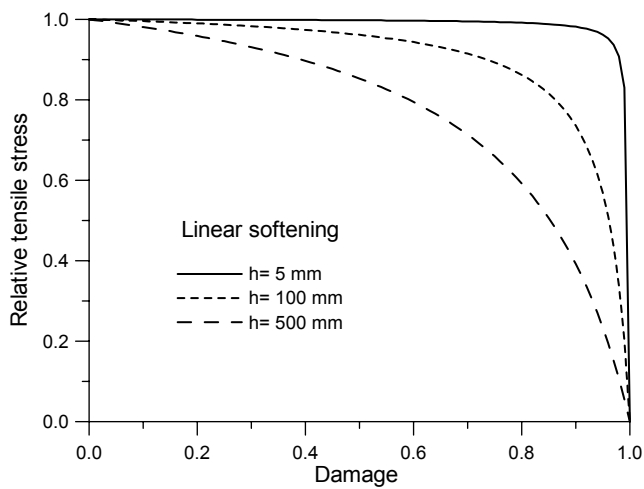

Figure 13. Dependency of the stress degradation on the damage for three different crack band widths.

In the present example the size of the finite elements (crack band width) was $h = 5$ mm which leads to a rather ductile stress-strain relationship that is almost elasto-plastic. Consequently, with such stress-strain law the model is not able to predict the crack path correctly. With a more realistic stress-strain relationship, i.e. by adopting the shape such that it is similar to an realistic stress-crack opening law, the predicted crack path becomes more objective. This shows that the present model is sensitive to the choice of the stress-strain curve, which is actually not a problem of the model but a limitation of the crack band approach. Namely, in a limit case ($h \to 0$) the solution yields to the plasticity solution.

The three curves shown in Figure 13 correspond to three different widths of the crack band (element size): $h = 5$, 100 and 500 mm. It can be seen that for small elements the stress does not reduce significantly until a very high level of damage. From the damage point of view the function is unrealistic, however, for the given element size the function is required to assure correct consumption of fracture energy (crack band method). There is an obvious need for the optimisation of the shape of the softening curve such that the stress is a realistic function of damage and at the same time that the consumption of energy is objective with respect to the element size.

## 6 CONCLUSIONS

A scalar damage model with loading surface in closed form is proposed without any explicit evolution law based on the equivalent strain concept. For the two-dimensional problem a forth order polynom is shown to be a realistic choice for the loading surface. It is shown that the formulation of the loading surface in the principal stress space has some advantages over the formulation in terms of stress invariants. The main advantage is transparency of the model and, what is of a great importance for the calibration of the model, it is easy to identify model parameters from experiments. The influence of different softening laws on the performance of the model is discussed. For a linear softening law and for a linear form of the loading surface, implicit damage evolution law is formulated in a closed form. It is demonstrated that in spite of its simplicity, the model can realistically predict the cracking of concrete for rather complex mixed mode fracture. If the crack band method is used as a localization limiter, then for small crack band width (small elements) the constitutive law tends to be almost elasto-plastic. This leads to unrealistic predictions and is in contradiction with quasi-brittle nature of fracture of concrete. More work is needed to further improve the proposed model.

## REFERENCES

Bažant, Z.P., Oh., B.H., 1983. Crack band theory for fracture of concrete. *Materials and Structures*, RILEM, 93(16): 155-177.

Etse, G., 1992. *Theoretical and numerical investigation of diffusive and localized damage in concrete*. Ph.D. Thesis, University of Karlsruhe. (in German).

Feenstra, P., 1993. *Computational aspects of biaxial stress in plain and reinforced concrete*. Ph.D. Thesis, Technical University of Delft.

Mazars, J., 1986. A description of micro- and macro-scale damage of concrete structures. *Journal of Engineering Fracture Mechanics,* 25: 729-737.

Meschke, G., Lackner, R., Mang, H., 1998. An anisotropic elastoplastic-damage model for plain concrete. *International Journal of Numerical Methods in Engineering*, 42: 703-727.

Ožbolt, J., Reinhardt, H.W., 2003. Numerical study of mixed mode fracture in concrete. *International Journal of Fracture*.Z.P., in press.

Nooru-Mohamed, M.B., 1992. *Mixed-mode fracture of concrete: an experimental approach*. Ph.D. Thesis, Technical University of Delft.

Pivonka, P., Ožbolt, J., Lackner, R., Mang, H., 2002. Comparative studies of 3D-constitutive models for concrete: application to mixed-mode fracture. *International Journal of Numerical Methods in Engineering*, (submitted).

Simo, J., 1992. Algorithms for static and dynamic multiplicative plasticity that preserve the classical return mapping schemes of the infinitesimal theory. *Computer Methods in Applied Mechanics and Engineering*, 99: 61-112.

Simo, J., Taylor, R., 1986. A return mapping algorithm for plane stress elastoplasticity. *International Journal of Numerical Methods in Engineering*, 29: 649-670.